# Linking Science and Industry: Influence of Scientific Research on Technological Innovation through Patent Citations

Pablo Dorta-González [a,*], Alejandro Rodríguez-Caro [b], María Isabel Dorta-González [c]

[a] Institute of Tourism and Sustainable Economic Development (TIDES), Campus de Tafira, University of Las Palmas de Gran Canaria, 35017 Las Palmas de Gran Canaria, Spain ORCID: http://orcid.org/0000-0003-0494-2903

[b] Department of Quantitative Methods in Economics and Management, Campus de Tafira, University of Las Palmas de Gran Canaria, 35017 Las Palmas de Gran Canaria, Spain; alejandro.rodriguez@ulpgc.es; ORCID: 0000-0002-8080-3094

[c] Department of Computer and Systems Engineering, Avenida Astrofísico Francisco Sánchez s/n, University of La Laguna, 38271 La Laguna, Spain; isadorta@ull.es; ORCID: 0000-0002-7217-9121

* Correspondence: pablo.dorta@ulpgc.es

**Abstract**

This study explores the connection between patent citations and scientific publications across six fields: Biochemistry, Genetics, Pharmacology, Engineering, Mathematics, and Physics. Analysing 117,590 papers from 2014 to 2023, the research emphasises how publication year, open access (OA) status, and discipline influence patent citations. Openly accessible papers, particularly those in hybrid OA journals or green OA repositories, are significantly more likely to be cited in patents, seven times more than those mentioned in blogs, and over twice as likely compared to older publications. However, papers with policy-related references are less frequently cited, indicating that patents may prioritise commercially viable innovations over those addressing societal challenges. Disciplinary differences reveal distinct innovation patterns across sectors. While academic visibility via blogs or platforms like Mendeley increases within scholarly circles, these have limited impact on patent citations. The study also finds that increased funding, possibly tied to applied research trends and fully open access journals, negatively affects patent citations. Social media presence and the number of authors have minimal impact. These findings highlight the complex factors shaping the integration of scientific research into technological innovations.

**Introduction**

The importance of scientific research goes beyond just academic circles. Research can greatly affect society by impacting various fields like education, culture, and the economy (Wilsdon et al. 2015). This wider societal impact can be measured quantitatively using altmetrics, a set of metrics that complement traditional citation analysis (NISO 2016).

The emergence of digital academic communication has transformed how we evaluate the social influence of research. This change promotes a broader strategy that takes into account a wider variety of research results and creative forms of communication (Bornmann 2013; de Rijcke et al. 2016; Bornmann & Haunschild 2019). An excellent illustration is the UK's Research Excellence Framework (REF 2021), which includes evaluating the influence of studies beyond the academic world. In particular, 25% of the REF assessment focuses on aspects such as influence on public policy, economic and social contributions, and improvements in health, the environment, and general well-being (Khazragui & Hudson 2015).

Research on altmetrics before 2015 primarily concentrated on online mentions and their relationship with citations, implying that all mentions carry equal social influence. Nevertheless, this was without a theoretical basis (Ravenscroft et al. 2017). The scholarly assessment literature distinguishes between the scientific impact within academia and the broader societal impact according to Spaapen & van Drooge (2011), Joly et al. (2015), and Morton (2015). Despite the initial excitement about altmetrics as a way to gauge societal impact due to the emphasis of funding agencies, the field now acknowledges the necessity for a new strategy. Recent research by Haustein et al. (2016), Robinson-García et al. (2018), Wouters et al. (2019), Costas et al. (2021), Dorta-González (2023), Alperin et al. (2024), and Dorta-González et al. (2024) suggests using altmetrics to gauge science-society interactions and knowledge sharing, rather than focusing solely on direct impacts.

It is important to investigate how patents are utilised in evaluating research institutions as the social influence of scientific research becomes more prominent. Such an exploration can be carried out by examining the connection between patents and research results, as shown by Chowdhury, Koya, & Philipson (2016) in the UK's Research Excellence Framework. This analysis focuses on a particular aspect of societal influence, specifically the consequences of technology, which are frequently connected to economic importance. Patents function as official papers that describe and defend technological advancements. Similar to scientific articles, patent documents contain citations. The main focus of these citations is on past patents for demonstrating originality in light of current technology, with some attention given to non-patent materials like scientific articles. Nonetheless, while authors are solely accountable for citations in scholarly articles, citations in patents can be contributed by both inventors and patent examiners (van Raan 2017a; Velayos-Ortega & López-Carreño 2023).

The disconnect between scientific knowledge creation and its incorporation into patents persists, and the impact of openly accessible scientific research on technological innovations is not well understood (Jahn et al. 2022). This study aims to close this divide by analysing how inventors and patent examiners cite sources. Our goal is to determine the most successful communication methods by analysing how these stakeholders mention research papers.

To achieve this goal, our study examines patent citations in various technological-related fields, including Biochemistry, Genetics, Pharmacology, Engineering, Mathematics, and Physics, categorized according to Lens.org's Field of Research (FoR). Using data extracted from journal articles from 2014 to 2023, and using Altmetric.com to query associated DOIs, we identified a corpus of 117,590 research papers. Our analysis includes different access types to these publications, as well as factors such as policy citations, which signal relevance to real-world challenges. We examine the potential impact of funding sources and the publication year, reflecting the lag in the integration of knowledge dissemination into patent citations. In addition, we examine scholarly metrics, including blog mentions, scholarly citations, and readership on platforms such as Mendeley, to measure visibility in academic circles. We also examine the influence of various social media platforms in the dissemination of research results.

**Patent citations to scientific literature**

Narin and his team carried out groundbreaking studies on patent citations to academic papers, as demonstrated by their influential publications (Carpenter, Cooper, & Narin 1980; Carpenter & Narin 1983; Narin & Noma 1985). They viewed the quantity of scientific citations as an indicator of the degree of scientific intensity in technological fields. Further studies conducted by Narin showed that around 75% of references in US industry patents were from public scientific sources such as academic and governmental institutions, while only 33% were written by industrial researchers. The scientific sources showed a strong tendency to cite papers from leading-edge contemporary science, especially fundamental research in influential journals, often associated with prestigious research institutions and well-supported by organizations like the US National Institutes of Health (NIH) and the US National Science Foundation (NSF) (Narin, Hamilton, & Olivastro 1997). This collection of work highlights the important role that a strong public research system plays in promoting technological progress and economic success.

Important prior contributions involve the initial investigation in merging patent data with bibliometric frameworks (Coward & Franklin 1989), which identified that a significant number of non-patent references in patents granted in the Netherlands during the 1980s were references to journal articles, mainly from journals listed in SCI-WoS (van Vianen, Moed, & van Raan 1990). This sentence further delved into research on patent-citation analysis, examining the use of scientifically important knowledge in both global and local contexts (Tijssen, Buter, & van Leeuwen 2000; Tijssen 2001).

According to Verbeek et al. (2002), the distribution of non-patent references in patents is imbalanced, as the majority of patents have no references while only a small number reference multiple sources, showcasing the connection between science and technology. Most of the references that were not patents were scientific references (Callaert et al. 2006). Furthermore, it was noted that a specific group of scientific journals primarily provided scientific references. Van Looy et al. (2006) showed that there is a strong link between a country's technological advancement and its scientific development, particularly when this knowledge is spread out among various businesses and research organizations. In new and growing technology fields, patents were more likely to include scientific references, showing a stronger reliance on scientific knowledge in innovative technological advances (van Looy, Magerman, & Debackere 2007).

Scientific data cited in patents frequently lacks consistency and standardization, as noted by Guerrero-Bote et al. (2021), resulting in multiple references that point to the same publication. A thorough integration of patent databases like PATSTAT is essential in order to accurately link scientific references with publications included in databases. Various data analysis methods are required to efficiently pair patent publications with data. Magerman, van Looy, & Song (2010) investigate text-mining methods to find similarities between patents and research articles. They advise not to view scientific citations as clear signs of a direct transfer of knowledge from research findings to patented technology. The way patents with scientific references are interpreted is greatly affected by their source, whether it's a university, public research institute, or company. Moreover, significant differences can be observed among nations, probably related to their scientific capacities. Moreover, differences in the usage of scientific citations in patent systems are significant (Callaert, Grouwels, & van Looy 2012), as the US Patent and Trademark Office (USPTO) requires more scientific references than the European Patent Office (EPO) or the World Intellectual Property Organization (WIPO).

Citing patents that refer to scientific literature provides important information about how technological advancements are spreading worldwide. Ribeiro et al. (2014) examined 167,000 patents registered by the USPTO in 2009, analysing the referenced papers to identify the international scientific impact of technology. Their research examines the complex relationships among multinational corporations, universities, and other companies on a global scale, highlighting the detailed dynamics of international technological partnerships.

Some scientific references are used for general background information rather than serving as crucial sources of inspiration for underlying research. In many instances, inventors view scientific references as unimportant or insignificant (Callaert, Pellens, & van Looy 2014). This difference in perception is frequently linked to variations in scientific references provided by inventors and examiners. According to Bryan et al. (2020), there is a notable discrepancy in how academic articles are cited in different parts of the patent. Specifically, only 24% of the academic article citations that appear on the front page of the patent also appear in the main text, suggesting that most references highlighted on the front page are not discussed or elaborated upon in detail in the document. Conversely, only 31% of citations included within the body of the text are present on the front page, implying that most references detailed in the text are not emphasised on the front page. In summary, these percentages reflect a difference in the focus or importance given to citations depending on their location within the patent.

According to Park et al. (2017), the way applicants and examiners cite patents can offer a glimpse into their technological worth. A greater number of citations from examiners is noticeable in applications from foreign individuals at the USPTO (Criscuolo & Verspagen 2008; Alcácer, Gittelman, & Sampat 2009). Citations from examiners show a stronger connection to existing knowledge than citations from applicants, particularly about patent claims (Chen 2017). Nevertheless, references provided by examiners come with their own set of difficulties. Wada (2016) points out barriers that examiners face when searching for prior art, showing that geographical distance reduces the chances of finding previous patents. Although this mainly applies to citing previous patents, equivalent challenges can also occur when citing scientific literature.

The level of scientific citations in patents and the likelihood of a publication receiving citations are influenced by several factors, including the contributions of inventors and examiners, as well as the characteristics of patent offices, companies, and technology sectors. Additionally, citations in patents to non-patent literature, largely consisting of scientific references, are not evenly spread out (Squicciarini et al. 2013). Scientific citations generally serve as a link between science and technology, impacting the depth of scientific knowledge in technological fields and the national relationship between science and technology (Han & Magee 2018).

The number of citations a scientific paper receives in patents is influenced by the time that elapses between its publication and its citation. This phenomenon, known as the 'sleeping beauty' effect, describes how some papers remain relatively unknown for a long time before being 'awakened' and cited in patents (van Raan 2017b; van Raan & Winnink 2018, 2019).

Based on the literature revision (van Raan 2017a; Velayos-Ortega & López-Carreño 2023), the following findings are identified. The majority of references in patents that are not patents themselves are scientific references, mainly drawn from a small number of journals. These scientific sources show a significant preference for scientific projects at the national and public levels. An unequal distribution of scientific references is observable in patents, with new fields typically featuring a higher number of scientific references. Nonetheless, scientific citations might not always clearly show the flow of knowledge and may not necessarily be the main references for the research being discussed. Moreover, scientific citations can be created by inventors or examiners, with examiners having a notable impact on their incorporation. Moreover, the quantity of scientific citations fluctuates based on the specific area of technology and its level of advancement.

**Methods**

*Data*

The paper aims to examine patent citations of scientific articles within various technological domains. For it, we used two data sources. The Lens (Lens.org) and Altmetric (Altmetric.com) databases provide complementary approaches to tracking citations in patents to scientific publications.

The Lens offers a dual functionality, enabling users to search for patent records from various offices, including the USPTO, EPO, and WIPO, and to explore the "References Cited" section, where citations to scientific literature are listed. Users can either search for a patent to see which scientific publications it cites or search for a scientific publication to find all patents that reference it. This bidirectional search capability facilitates the exploration of the interplay between patent innovation and scientific research. On the other hand, Altmetric focuses on the impact and attention received by scientific publications. It tracks mentions and citations from diverse sources, including patents. In Altmetric, users start with a scientific publication and can view the number of times it has been cited in patents, among other impact metrics. These citations contribute to the publication's overall Altmetric Attention Score. Both databases provide valuable insights into how scientific research influences technological developments and patent filings, although The Lens focuses more on the patent side, while Altmetric centres on the attention and reach of the scientific work.

We chose six disciplines from the Field of Research (FoR) classification in The Lens. The FoR classification is a document-level categorization employed in databases like Dimensions, Altmetric, and The Lens. To focus on areas with a substantial technological impact, we selected disciplines based on their demonstrated capacity to drive innovation and their prominence in both academic research and patent activity. Specifically, three disciplines were chosen from the Life Sciences – Biochemistry, Genetics, and Pharmacology – because of their critical role in advancing biomedical technologies and translating research into practical applications. Likewise, three disciplines were selected from the Physical Sciences – Engineering, Mathematics, and Physics – due to their essential contributions to technological advancements and industrial processes. These fields were chosen on the basis that they offer a representative cross-section of sectors where the interplay between scientific research and technological development is particularly pronounced, thereby strengthening the rigour of our methodological approach.

The study focused on journal articles as the documentary type, and the publication timeframe was restricted from 2014 to 2023. We acknowledge that including publications from 2014 to 2023 may introduce a citation lag, particularly for the most recent articles. However, our model incorporates time as an explanatory variable, which enables us to compare the other effects with those associated with the passage of time. This approach allows us to explicitly account for the delay in accumulating patent citations, thereby mitigating the potential bias from including newer publications.

Following these parameters, a set of 159,522 papers containing at least one patent citation was retrieved from the The Lens database. The scholar search query used in The Lens was as follows; field_of_study: (Biochemistry OR Genetics OR Pharmacology OR Engineering OR Mathematics OR Physics) AND publication_type: ("journal article") AND is_referenced_by_patent: true AND (year_published:2014 OR year_published:2015 OR year_published:2016 OR year_published:2017 OR year_published:2018 OR year_published:2019 OR year_published:2020 OR year_published:2021 OR year_published:2022 OR year_published:2023)

Subsequently, the DOI was queried on Altmetric database. Under these conditions, 117,590 papers were identified in Altmetric. Data extraction was performed on February 4, 2024.

*Variables (Scientific citations vs. Patent citations)*

Patent citations differ from other types of citations in several ways (van Raan 2017a). Firstly, their purpose diverges: patent citations constrain what an applicant can claim as new or novel by referencing prior art, while scientific citations acknowledge sources and contribute to existing knowledge. Secondly, the examination process varies: patent citations are also added by examiners to identify relevant prior art, while scientific citations are only included by authors. Thirdly, incentives for citation inclusion vary: patent applicants aim to minimize citations to avoid limiting their claims, whereas scientists cite broadly to establish the foundations of their work. Finally, interpretation diverges: the number of patent citations indicates technological importance, whereas scientific citations measure academic influence.

*Methods*

We opted for Poisson regression over negative binomial regression based on considerations of parsimony, robustness of inference, and interpretability. While the negative binomial estimator can offer greater efficiency in certain scenarios, empirical evidence suggests that such efficiency gains are often marginal in practice (Cameron & Trivedi 2022). Our analysis focuses on modelling the conditional mean – not conditional probabilities – and Poisson regression remains a valid choice for this objective, provided the conditional mean is correctly specified. As emphasized by Cameron and Trivedi (2022), Poisson models retain their appropriateness for conditional mean estimation even in the presence of overdispersion, provided robust standard errors are employed to address inference challenges. By adopting this approach, we preserved the model's simplicity while ensuring the validity of our results, thereby enhancing the clarity and interpretability of our findings for applied audiences. To further validate our choice, we conducted post-estimation tests for overdispersion (e.g., likelihood-ratio tests) and confirmed that the use of robust standard errors adequately mitigates potential concerns. This strategy aligns with recommendations in econometric literature for settings where interpretability and parsimony are prioritized over distributional assumptions.

While the reported pseudo-$R^2$ (0.1025) may appear low compared to linear regression benchmarks, this metric is inherently limited in its interpretability for count models like Poisson regression. As emphasized in Cameron & Trivedi (2022), pseudo-$R^2$ values in such frameworks do not measure explained variance as in linear models and often yield modest values even for well-specified models. This is particularly true in large samples like ours – with more than one hundred thousand observations –, where even minor deviations between predicted and observed counts can depress metrics like the squared correlation between predictions and outcomes.

Our primary focus lies in inference – estimating consistent coefficients for explanatory variables – rather than maximizing predictive accuracy. For this purpose, Poisson regression remains valid and widely recommended when robust standard errors are employed to address potential overdispersion (Wooldridge 2010). We conducted post-hoc goodness-of-fit assessments, including residual dispersion tests and visual inspections of predicted vs. observed counts. These diagnostics confirmed no systematic misfit or heteroscedasticity patterns that would undermine our inferential goals.

We utilized a substantial dataset of 112,969 observations following the removal of incomplete records. The Chi-squared test was used to evaluate the statistical importance of the correlation between the predictor variables and the dependent variable, with a *p-value* of less than $10^{-3}$ suggesting a strong connection.

To clarify how our predictor variables influence the outcome, we calculated average marginal effects using the delta method. In our study, '$y$' represents the dependent variable – the number of patent citations – while '$x$' denotes each independent explanatory variable included in the regression. Calculating '*dy/dx*', the first derivative of the function $y = f(x)$, illustrates how a one-unit change in a specific independent variable is expected to impact the average number of patent citations while keeping all other factors constant. For example, if $x$ corresponds to the number of mentions in blogs and the *dy/dx* value is 0.07, this suggests that a one-unit increase in blog mentions is associated with an expected increase of 0.07 patent citations, ceteris paribus. Similarly, ten blog mentions would correspond to an expected increase of 0.7 patent citations, all else being equal.

This method provides a clearer understanding of the interrelationships between variables and offers insights into how changes in each factor affect the outcome of interest. By examining *dy/dx* for each independent variable, readers can discern both the magnitude and direction of the effects, improving the interpretation of the results and informing decision-making processes.

**Results and discussion**

Table 1 gives a summary of research papers in the sample from various fields and time periods. It is evident that the number of research papers cited by patents declines as the year of publication rises, possibly because of the delay between article publication and patent citation – as well as the gap between patent application and publication –.

Significantly, there has been a noticeable rise in the overall count of articles referenced in patents published in 2022 when compared to past years. This increase in articles cited in patents, soaring from 5,000 in 2021 to 13,154 in 2022, is remarkable. It is not limited to specific fields but is widespread across various disciplines, indicating a broad impact. For instance, Biochemistry experienced a remarkable rise from 1,212 articles in 2021 cited in patents to 3,130 in 2022. Similarly, disciplines such as Engineering, Genetics, Mathematics, and Physics also experienced significant increases in cited articles during this period. The surge in Engineering citations is particularly striking, with the count escalating from 422 to 3,247 articles cited in patents in just one year. Conversely, Pharmacology saw a decline in cited articles from 1,075 to 802. This is a notable contrast to the overall trend of increasing patent citations across all disciplines. The decline in cited articles in Pharmacology may be due to shifts in research focus or methodology in referencing scientific evidence within the field during the period under study, leading to fewer citations in patents.

The generalized increase in publications in 2022 cited in patents can be attributed to several factors related to the COVID-19 pandemic. Firstly, the urgent need for research into the virus and its societal impact is likely to have led to an increase in publications across a range of disciplines. In addition, the increase in patents related to the pandemic and the subsequent referencing of scientific articles within these patents are likely to have contributed to this trend.

Table 2 highlights the important role of funding in determining the accessibility of articles cited in patents. It indicates that around two-thirds of the referenced articles in patents have been supported financially, with notable variations across different fields of study. In Engineering, for example, only 53% of cited articles have received funding, whereas in Genetics the figure is as high as 80%. This disparity is also reflected in the type of access provided, with a higher percentage of closed access (55%) in Engineering and a lower percentage (17%) in Genetics.

In Table 2, Genetics emerges as the most well-funded discipline, with 80% of its articles disclosing funding sources. Of these, 42% are published in fully open access (gold) journals. Biochemistry and physics follow Genetics with 77% and 75% of funded articles respectively. About one in four articles in these disciplines opt for publication in fully open access (gold) journals (26%). By contrast, Engineering stands out as the least funded discipline, with only 53% of its articles indicating funding sources, and of these, only 17% opting for fully OA journals.

This observation leads to the conclusion that disciplines with higher proportions of funding tend to prefer to publish in fully open access (gold) journals. In other words, higher levels of funding correlate with a higher likelihood of publishing in gold OA journals. Hence, funding is essential in deciding the degree of access to open articles (Dorta-González & Dorta-González 2023a, 2023b). This significant finding emphasizes the crucial role of sufficient financial support in guaranteeing the widespread dissemination and free access of research to the public. The expenses linked to publishing, such as article processing charges (APCs), can pose a significant obstacle to certain types of open access, especially for researchers from institutions with limited funding or lacking access to institutional funding. Therefore, the higher the level of funding, the more likely it is that research will be published in some form of open access, thereby increasing its visibility and accessibility to a wider audience.

Table 3 presents descriptive statistics for various variables, which show highly skewed distributions. The average number of patent citations for research articles in the sample is 1.89, with a standard deviation of 5.73 in the Altmetric database. This distribution is highly skewed, with a minimum of 0 citations and a maximum of 441. The average number of mentions in news articles is 1.50, with a standard deviation of 14.81, ranging from 0 to 2,048. Mentions in blog posts have an average of 0.23. Across social media platforms, the average number of mentions is 14.02 on X/Twitter and 0.42 on Facebook. Wikipedia mentions an average of 0.19 times the articles in the sample. Mendeley averages 96.44 per article, reflecting a wide range of readership. On average, scientific citations stand at 76.19, with a standard deviation of 316.83 and a range from 0 to 56,303. Lastly, the number of authors per article varies widely, with an average of 6.66 and a standard deviation of 12.66, ranging from 1 to 2,082.

The correlation matrix (Table 4) shows that patent citations are a distinct metric, with some overlap with conventional academic impact (scientific citations) and researcher interest (Mendeley readers). However, they show less direct correlation with online attention, policy references, or the number of authors involved in the research. Most of the values in the patent citations column are close to 0, indicating very weak correlations with other variables such as news mentions, social media mentions, policy citations, and the number of authors in the research article. An exception is the weak positive correlation with scientific citations (0.16), inferring that articles with higher scientific citations (presumably reflecting high-impact research) tend to receive slightly more patent citations. Furthermore, there is a moderate positive correlation of 0.27 with Mendeley readers, indicating that research articles which garner significant attention on Mendeley, a tool used for managing references by researchers, are more prone to being referenced in patents. On the other hand, the slight negative association with the year of publication (-0.10) suggests that articles published in the past have had a longer period to gather patent citations compared to more recent articles, despite the correlation being weak.

The Poisson regression model in Table 5, based on 112,969 data points, displays the below statistics using the maximum likelihood fitting approach. The *p-value* of the Chi-square test is below $10^{-3}$, showing a significant link between predictor variables and the dependent variable. The pseudo-$R^2$ value of 0.1025 is not high, but it does allow conclusions to be drawn about the effect of each explanatory variable when the others are held constant (ceteris paribus). In addition, average marginal effects are reported using the delta method. The *dy/dx* shows how the expected value of patent citations changes with a one-unit change in the independent variable while keeping all other variables constant. This method allows for the determination of both the size and the orientation of the impacts.

The results of the regression model presented in Table 5 reveal numerous statistically significant findings that provide insights into the correlation between different variables and patent citations. The blog post mentions and patent citations have a positive correlation, with a coefficient of 0.0379 and a *p-value* below $10^{-3}$. This indicates an increase in blog mentions is associated with an increase in patent citations when everything else is held constant. In contrast, policy citations, scientific citations, and funding have negative effects on patent citations. Specifically, policy citations have a coefficient of -0.1465 and a *p-value* of 0.027, indicating a significant negative effect on patent citations. Similarly, scientific citations have a coefficient of -0.0002 and a *p-value* less than $10^{-3}$, showing a statistically significant negative effect on patent citations. Funding also has a significant negative effect, with a coefficient of -0.0869 and a *p-value* less than $10^{-3}$, indicating that funded research is associated with fewer patent citations. In each case, holding all other variables constant. This relates to the negative impact of the gold OA model and the conclusion drawn from Table 2 regarding the disciplines that show a preference for fully OA (gold) journals at higher funding levels. Basically, there is a connection indicating that increased levels of funding are linked to a greater chance of being published in gold open access journals.

Conversely, mentions on social media platforms – including X/Twitter and Facebook – do not show a statistically significant impact on patent citations, as demonstrated by the *p-value*. Likewise, references on Wikipedia have a negligible impact. There is no substantial impact of the number of authors on patent citations, according to a *p-value* of 0.226. Yet, its importance may be clear in academic and social settings (Dorta-González & Dorta-González 2022).

Furthermore, the analysis shows a significant negative impact of the publication year on patent citations, with coefficients increasing consistently in absolute terms from 2015 to 2023. This indicates that more recent articles generally receive fewer patent citations than older ones. The trend can be attributed to the delay in publishing articles before applying for patents and the delay in publishing patents after applying for them. Additionally, there are notable statistical variances between open and closed access articles in terms of their citation in patents, as well as across different fields in comparison to Genetics.

The size of the effects can be assessed based on the magnitude of the marginal effects (*dy/dx*) reported for each variable, other variables held constant. The analysis shows a range of effects on patent citations, from large to small, depending on the variable in question.

*Large effects*

The marginal effects for open access compared to closed access show different sizes depending on the type of open access, ranging from -0.3887 to 0.5095. Publication access types can be ranked according to their likelihood of being cited in patents. Research papers from well-known subscription-based journals that do not offer open access are cited more frequently than those from completely open access journals. However, articles in hybrid OA journals with author or institution-paid APCs are more likely to be cited in patents than other articles in the journal. Similarly, academic papers saved in a specialized or institutional open access repository (green OA) have a higher likelihood of being cited in patents, just slightly edging out the hybrid method.

On the other hand, the marginal effect of policy citations is -0.2755, indicating that for every additional citation in policy documents, patent citations decrease by about 0.27 units, which is a relatively large negative effect. The marginal effects of the publication year indicate a significant decrease in patent citations over time, ranging from -0.1879 to -1.8425. This indicates a large and consistent negative effect of publication year on patent citations. It means that the probability of being cited in patents increases by about 0.18 with each passing year. This phenomenon can be clarified by two significant delays in time. There is frequently a delay between when a research paper is published and when it is referenced in a patent filing. This is because it requires time for research findings to be acknowledged, comprehended, and subsequently utilized in the creation of innovative inventions and technologies. A delay also occurs between the initial application of a patent and its eventual release for public access. The field also plays a role in the time it takes for research to be cited in patents. While in certain fields, such as virology or pharmaceuticals, scientific papers may be cited in patents more quickly due to the immediacy of their applications, other fields, such as fundamental physics, may require significantly longer periods before their scientific findings are used for technological purposes. The variation across fields underlines the importance of taking into account both the nature of the discovery and the broader societal context when assessing the timeline from scientific research to technological impact.

*Moderate effects*

The marginal effect of blog mentions is 0.0713, indicating that each additional mention in a blog post corresponds to an increase of approximately 0.07 patent citations, denoting a moderately positive effect. Conversely, the marginal effect of funding is -0.1635, indicating that the probability of being cited in a patent decreases by about 0.16 units for funded research compared to unfunded ones. It's worth noting that increased funding levels are correlated with a stronger tendency to publish in gold open access journals, which in turn has a negative impact on patent citations.

*Small effects*

The marginal effect of readers on Mendeley is 0.0012, signifying that for every additional thousand readers on Mendeley, patent citations increase by approximately 1.2 units, which is a small positive effect. Conversely, the marginal effect of scholarly citations is -0.0003, indicating that for every additional thousand scholarly citations, citations in patents decrease by about 0.3 units, a very small negative effect – although statistically significant –.

*Different effects*

The marginal effect for different disciplines compared to Genetics shows different sizes depending on the field, ranging from -0.4308 to -1.0926. These results suggest that the impact of the discipline on patent citations is important. According to the marginal effect on patent citations, the research in the disciplines analyzed can be ranked from highest to lowest probability of being cited in patents: Genetics, Pharmacology, Engineering, Biochemistry, Physics, and Mathematics.

In summary, the analysis reveals that variables such as the access type, the number of citations in policies (although inversely) – indicative of practical application to societal issues –, the publication year, and the discipline, exert considerable influence on the likelihood of being cited in patents. Conversely, scientific variables like blog mentions, readership on Mendeley, and funding (although inversely) exhibit smaller effects. So, making scientific publications openly accessible significantly increases their use in patent applications, especially when they are published in recognized subscription-based journals where the authors, often supported by their institutions, have financed the publication costs (hybrid OA) or via open access repositories (green OA), in both cases about seven times more than for a mere mention in a blog, and more than twice as much as for a passing year.

Publications that are openly accessible, particularly those from recognised, institutionally sponsored subscription-based journals or through open access repositories, demonstrate markedly higher patent citation rates compared to informal blog mentions or temporal factors. The observed negative association between funding levels and patent citations may partially stem from the reliance on article processing charges linked to fully open access (gold) journals, though this analysis did not account for potential confounding variables such as variations in journal quality or disciplinary norms. By contrast, hybrid models – supported by institutional-publisher transformative agreements – appear to attenuate this effect, potentially reflecting the influence of institutional policies on dissemination practices. While traditional subscription-based journals, often perceived as more prestigious, likely shape citation behaviours, the discussion of open access mechanisms requires careful interpretation, as the interplay between publication formats, institutional mandates, and inherent research quality remains insufficiently explored.

Another factor could be the shift towards funding applied research rather than basic research, as the latter often leads to basic knowledge that is more frequently cited in patents. A redirection of research funding towards applied-driven projects may consequently reduce patent citation rates, as they tend to rely on basic research results. This is also related to the negative correlation observed with policy citations. Disciplinary differences may also play a role, with some fields being more prone to produce patent-cited papers. Any imbalance in the distribution of funding between disciplines could distort the results, as can be seen in Table 2.

However, this study is not without limitations that could be addressed in future research. First, the study relies on a specific dataset of patent citations that may not capture the full range of citations or may be subject to biases inherent in patent databases. In addition, limitations in the availability or quality of data for certain variables may affect the robustness of our findings.

Additionally, time plays a crucial role in the interaction between science and technology, particularly in how quickly scientific information is transferred to the patenting process (Mehta, Rysman, & Simcoe 2010). The time lag between a paper being published and cited in a patent varies greatly across technology sectors, impacting the age of scientific references. In newly developing areas, the delay is usually brief. As an example, Finardi (2011) notes a delay of three to four years in the field of nanotechnology. Nevertheless, several studies indicate significantly extended time delays. It is common for fundamental scientific discoveries to take more than one generation before being applied to new technologies (Sherwin & Isenson 1967). Grant, Green, and Mason (2003) mention that there exists a 20-year gap between advancements in neonatal intensive care and fundamental research.

In this respect, our analysis focuses on a specific period of ten years, which may not be sufficient to analyze a phenomenon with long delays between the publication of a paper and its first citations in patents. Trends observed within this period may not necessarily generalize to other periods due to evolving publication and patenting practices. Future research should consider longer longitudinal analyses to assess how citation patterns evolve.

Although efforts were made to control for disciplinary differences, caution should be exercised when extrapolating these findings to other research areas or contexts. Differences in disciplinary norms, research funding structures, and patenting behaviour may limit the generalisability of our findings. Furthermore, the number of scientific citations in patents depends on the progress of a technology field. Multiple researches examined the analysis of patent citations in growing and evolving areas like nanotechnology (Hu et al. 2007; Meyer 2000, 2001) and genetic engineering research (Lo 2010). A quickly changing technological sector leans on more up-to-date scientific information than an established sector.

Although the study identifies associations between variables such as type of access, funding, and patent citation rates, it remains difficult to establish causality. Factors not included in our analysis, such as publication quality or researcher reputation, may confound these relationships (Velayos-Ortega & López-Carreño 2023).

**Conclusions**

Our analysis reveals significant factors influencing the likelihood of scientific publications being cited in patents. Access type, policy citations, publication year, and disciplinary domain have significant effects, while variables such as blog mentions, readership on Mendeley, and funding have more modest effects.

In particular, openly accessible publications – notably those originating from recognised, institutionally sponsored, subscription-based journals or available via open access repositories – exhibit significantly higher patent citation rates than mere blog mentions or the simple passage of time.

The observed negative correlation between funding and patent citation rates might be partly attributable to the reliance on funding associated with publishing in fully open access (gold) journals. In contrast, hybrid models, supported by agreements between publishers and institutions, appear to mitigate this effect. It is also plausible that traditional subscription-based journals, generally regarded as more prestigious, exert an influence on citation patterns. However, the differential impact of various open access models (gold, green, hybrid) remains open to debate, as the underlying mechanisms are not convincingly elucidated; confounding factors such as journal quality and institutional policies have not been adequately considered.

In addition, a shift towards funding applied research rather than basic research may contribute to lower patent citation rates, as the latter tends to provide the basic knowledge often cited in patents. This shift correlates with the negative effect observed for policy citations, reflecting a possible reorientation of research funding towards practical applications.

It is important to stress that these associations are purely correlational, and no causal claims can be inferred from the current data. Further investigation is necessary to substantiate these observations and to clarify the underlying mechanisms at play.

Disciplinary differences further underline the complexity of patent citation dynamics, with certain fields showing a higher propensity for patent-cited publications. Imbalances in the distribution of funding across disciplines could distort these dynamics, highlighting the need for equitable resource allocation.

Overall, our findings shed light on the multifaceted interplay between funding, access modes, publication dynamics, and disciplinary factors in shaping patent citation patterns, providing valuable insights for policymakers, researchers, and funding agencies alike.

It is important to acknowledge that several potential confounding variables have not been fully accounted for in this study. One key limitation is the possibility of selection bias in the patent database used. Certain patents may be more likely to be included or cited due to factors such as geographical location, industry focus, or filing strategies, which could skew the results. Additionally, differences in citation practices across fields have not been considered; certain disciplines may have inherently higher or lower citation rates due to variations in research dissemination, collaboration norms, or the time required for technological impact to manifest.

Moreover, influence of the institutional type and prestige is likely significant, as more renowned institutions may benefit from enhanced research infrastructure, visibility, and collaborative networks. These factors can affect both the quality of the research output and its subsequent translation into patent citations. Additionally, the effects of international collaboration have not been explicitly incorporated; such partnerships often foster greater exchange of ideas and access to diverse resources, which may, in turn, influence citation patterns.

Another consideration is the variation in practices among different patent offices, such as the USPTO, EPO, and WIPO. Each of these institutions has its own set of citation practices and examination procedures, potentially leading to systematic differences in how patent citations are recorded and valued. Addressing these factors in future research would provide a more comprehensive understanding of the dynamics underlying the relationship between academic output and patent citations, thereby enhancing the robustness and generalisability of our findings.

Table 1: Distribution of research articles in The Lens by discipline and year

| Discipline | Publication Year | | | | | | | | | | |
|---|---|---|---|---|---|---|---|---|---|---|---|
| | 2014 | 2015 | 2016 | 2017 | 2018 | 2019 | 2020 | 2021 | 2022 | 2023 | Total |
| Biochemistry | 9174 | 8252 | 7133 | 5827 | 2285 | 1944 | 1413 | 1212 | 3130 | 532 | 40,902 |
| Engineering | 5586 | 5597 | 5621 | 5096 | 1035 | 343 | 204 | 422 | 3247 | 736 | 27,887 |
| Genetics | 5789 | 5484 | 4470 | 3485 | 1306 | 1012 | 774 | 799 | 2077 | 324 | 25,520 |
| Mathematics | 2484 | 2264 | 2098 | 1766 | 1015 | 716 | 525 | 415 | 1299 | 386 | 12,968 |
| Pharmacology | 5514 | 5209 | 4826 | 4080 | 2375 | 1992 | 1872 | 1075 | 802 | 100 | 27,845 |
| Physics | 3724 | 3505 | 3260 | 2931 | 2684 | 2370 | 1637 | 1077 | 2599 | 613 | 24,400 |
| **Total** | **32,271** | **30,311** | **27,408** | **23,185** | **10,700** | **8377** | **6425** | **5000** | **13,154** | **2691** | **159,522** |

Table 2: Distribution of research articles in Altmetric by discipline, access type, and funding

| | Access Type | | | | | | | | | | | | | | | | All | | |
| | Gold | | | | Green | | | Hybrid | | | Bronze | | | Closed | | | | | |
| **Discipline** | **Funded** | **% of Total Gold** | **% of All Funded** | **Total** | **Funded** | **%** | **Total** | **Funded** | **%** | **Total** | **Funded** | **%** | **Total** | **Funded** | **%** | **Total** | **Funded** | **%** | **Total** |
|---|---|---|---|---|---|---|---|---|---|---|---|---|---|---|---|---|---|---|---|
| Biochemistry | 7126 | 78 | 26 | 9194 | 4832 | 88 | 5503 | 3886 | 84 | 4642 | 2932 | 76 | 3848 | 8931 | 70 | 12,680 | 27,707 | 77 | 35,867 |
| Engineering | 1732 | 46 | 17 | 3754 | 1599 | 62 | 2588 | 1082 | 76 | 1417 | 372 | 41 | 907 | 5447 | 52 | 10,462 | 10,232 | 53 | 19,128 |
| Genetics | 3897 | 82 | 42 | 4757 | 1356 | 89 | 1522 | 1485 | 85 | 1737 | 1304 | 77 | 1701 | 1294 | 66 | 1946 | 9336 | 80 | 11,663 |
| Mathematics | 845 | 58 | 16 | 1466 | 1339 | 68 | 1968 | 596 | 78 | 768 | 274 | 51 | 540 | 2087 | 58 | 3596 | 5141 | 62 | 8338 |
| Pharmacology | 4293 | 65 | 28 | 6586 | 2692 | 79 | 3392 | 1764 | 68 | 2610 | 2061 | 61 | 3405 | 4496 | 50 | 8905 | 15,306 | 61 | 24,898 |
| Physics | 3466 | 78 | 26 | 4418 | 3469 | 82 | 4213 | 1760 | 90 | 1964 | 666 | 67 | 999 | 3810 | 63 | 6055 | 13,171 | 75 | 17,649 |
| **Total** | **21,359** | **71** | **26** | **30,175** | **15,287** | **80** | **19,186** | **10,573** | **80** | **13,138** | **7609** | **67** | **11,400** | **26,065** | **60** | **43,644** | **80,893** | **69** | **11,7543** |

Table 3: Descriptive of quantitative variables in Altmetric (Obs. 117,590)

| Variable | Mean | Std. Dev. | Min | Max |
|---|---|---|---|---|
| Patent citations | 1.89 | 5.73 | 0 | 441 |
| News mentions | 1.50 | 14.81 | 0 | 2,048 |
| Blog mentions | 0.23 | 1.44 | 0 | 102 |
| Policy citations | 0.04 | 0.76 | 0 | 121 |
| X/Twitter mentions | 14.02 | 340.19 | 0 | 50,099 |
| Facebook mentions | 0.42 | 8.66 | 0 | 2,435 |
| Wikipedia mentions | 0.19 | 1.80 | 0 | 293 |
| Mendeley readers | 96.44 | 233.00 | 0 | 25,013 |
| Scientific citations | 76.19 | 316.83 | 0 | 56,303 |
| Num authors | 6.66 | 12.66 | 1 | 2,082 |

Table 4: Correlations between variables

| | Patent citations | News mentions | Blog mentions | Policy citations | X/Twitter mentions | Facebook mentions | Wikipedia mentions | Mendeley readers | Scientific citations | Num authors | Publication year |
|---|---|---|---|---|---|---|---|---|---|---|---|
| Patent citations | 1 | | | | | | | | | | |
| News mentions | 0.04 | 1 | | | | | | | | | |
| Blog mentions | 0.15 | 0.62 | 1 | | | | | | | | |
| Policy citations | 0.03 | 0.26 | 0.24 | 1 | | | | | | | |
| X/Twitter mentions | 0.02 | 0.45 | 0.27 | 0.10 | 1 | | | | | | |
| Facebook mentions | 0.01 | 0.09 | 0.14 | 0.03 | 0.06 | 1 | | | | | |
| Wikipedia mentions | 0.09 | 0.21 | 0.38 | 0.10 | 0.10 | 0.08 | 1 | | | | |
| Mendeley readers | 0.27 | 0.25 | 0.37 | 0.21 | 0.10 | 0.07 | 0.28 | 1 | | | |
| Scientific citations | 0.16 | 0.16 | 0.27 | 0.26 | 0.06 | 0.04 | 0.22 | 0.77 | 1 | | |
| Num authors | 0.02 | 0.07 | 0.13 | 0.01 | 0.03 | 0.02 | 0.16 | 0.06 | 0.07 | 1 | |
| Publication year | -0.10 | 0.05 | 0.01 | -0.01 | 0.04 | -0.01 | -0.03 | -0.10 | -0.08 | 0.04 | 1 |

Table 5: Robust Poisson regression model on patent citations (Obs. 112,969)

|  | Coef. | Std. Err. | z | P>z | Lower limit (95%) | Upper limit (95%) | dy/dx |
|---|---|---|---|---|---|---|---|
| News mentions | -0.0006 | 0.0009 | -0.63 | 0.530 | -0.0023 | 0.0012 | -0.0011 |
| Blog mentions | 0.0379 | 0.0099 | 3.84 | 0.000 | 0.0186 | 0.0572 | 0.0713 |
| Policy citations | -0.1465 | 0.0663 | -2.21 | 0.027 | -0.2763 | -0.0166 | -0.2755 |
| X/Twitter mentions | 0.0000 | 0.0000 | 0.93 | 0.353 | 0.0000 | 0.0001 | 0.0001 |
| Facebook mentions | -0.0012 | 0.0013 | -0.91 | 0.364 | -0.0038 | 0.0014 | -0.0022 |
| Wikipedia mentions | 0.0036 | 0.0077 | 0.47 | 0.639 | -0.0115 | 0.0187 | 0.0068 |
| Mendeley readers | 0.0006 | 0.0001 | 7.6 | 0.000 | 0.0005 | 0.0008 | 0.0012 |
| Scientific citations | -0.0002 | 0.0000 | -4.34 | 0.000 | -0.0003 | -0.0001 | -0.0003 |
| Funding | -0.0869 | 0.0162 | -5.38 | 0.000 | -0.1186 | -0.0552 | -0.1635 |
| Num authors | -0.0019 | 0.0016 | -1.21 | 0.226 | -0.0051 | 0.0012 | -0.0037 |
| Publication year (concerning 2014) | | | | | | | |
| 2015 | -0.0796 | 0.0271 | -2.94 | 0.003 | -0.1327 | -0.0265 | -0.1879 |
| 2016 | -0.2265 | 0.0269 | -8.42 | 0.000 | -0.2792 | -0.1738 | -0.4975 |
| 2017 | -0.3987 | 0.0246 | -16.18 | 0.000 | -0.4469 | -0.3504 | -0.8072 |
| 2018 | -0.5463 | 0.0281 | -19.44 | 0.000 | -0.6013 | -0.4912 | -1.0333 |
| 2019 | -0.6463 | 0.0334 | -19.32 | 0.000 | -0.7118 | -0.5807 | -1.1686 |
| 2020 | -0.7376 | 0.0321 | -22.99 | 0.000 | -0.8005 | -0.6747 | -1.2809 |
| 2021 | -0.8752 | 0.0285 | -30.74 | 0.000 | -0.9310 | -0.8194 | -1.4318 |
| 2022 | -1.1273 | 0.0256 | -44.04 | 0.000 | -1.1775 | -1.0772 | -1.6598 |
| 2023 | -1.3884 | 0.0388 | -35.8 | 0.000 | -1.4644 | -1.3124 | -1.8425 |
| Open Access (concerning closed) | | | | | | | |
| gold | -0.2395 | 0.0205 | -11.67 | 0.000 | -0.2798 | -0.1993 | -0.3887 |
| green | 0.2463 | 0.0294 | 8.39 | 0.000 | 0.1887 | 0.3038 | 0.5095 |
| hybrid | 0.2325 | 0.0276 | 8.44 | 0.000 | 0.1785 | 0.2865 | 0.4776 |
| bronze | 0.0837 | 0.0308 | 2.71 | 0.007 | 0.0232 | 0.1441 | 0.1592 |
| Discipline (concerning Genetics) | | | | | | | |
| Biochemistry | -0.3572 | 0.0387 | -9.23 | 0.000 | -0.4330 | -0.2813 | -0.7627 |
| Engineering | -0.3423 | 0.0408 | -8.39 | 0.000 | -0.4222 | -0.2624 | -0.7360 |
| Mathematics | -0.5626 | 0.0468 | -12.02 | 0.000 | -0.6544 | -0.4709 | -1.0926 |
| Pharmacology | -0.1859 | 0.0382 | -4.87 | 0.000 | -0.2607 | -0.1111 | -0.4308 |
| Physics | -0.4741 | 0.0439 | -10.79 | 0.000 | -0.5602 | -0.3879 | -0.9586 |
| Cons. | 1.1764 | 0.0404 | 29.14 | 0.000 | 1.0973 | 1.2556 | |